# A Dynamic 'Double Slit' Experiment in a Single Atom.

James Pursehouse[1], Andrew James Murray[1*], Jonas Wätzel[2] and Jamal Berakdar[2]

1. Photon Science Institute, School of Physics & Astronomy, University of Manchester, Manchester M13 9PL, UK
2. Martin-Luther-Universität Halle-Wittenberg, Institute of Physics, 06099 Halle/Saale, Germany

*Email: Andrew.Murray@manchester.ac.uk

**Abstract:**

A single-atom 'double-slit' experiment is realized by photo-ionizing Rubidium atoms using two independent low power lasers. The photoelectron wave of well-defined energy recedes to the continuum either from the 5P or 6P states in the same atom, resulting in two-path interference imaged in the far field using a photoelectron detector. Even though the lasers are independent and not phase locked, the transitions within the atom impart the phase relationship necessary for interference. The experiment is designed so that either 5P or 6P states are excited by one laser, before ionization by the second beam. The measurement cannot determine which excitation path is taken, resulting in interference in wave-vector space analogous to Young's double-slit studies. As the lasers are tunable in both frequency and intensity, the *individual* excitation-ionization pathways can be varied, allowing dynamic control of the interference term. Since the electron wave recedes in the Coulomb potential of the residual ion, a quantum model is used to capture the dynamics. Excellent agreement is found between theory and experiment.

Following Thomas Young's demonstration of the wave-nature of light through his 'double-slit' experiment around 1801, von Laue and others demonstrated matter-wave interference in the early 1900's. This included experiments by Davisson and Germer in 1927 that confirmed De Broglie's hypothesis of matter-waves for electrons. The first double-slit experiment using electrons was conducted in 1961 [1] and was later demonstrated for $C_{60}$ [2] and larger molecules [3]. Interference is also observed in atomic processes, including Fano resonances [4] and production of quantum vortices in ionization studies [5-7]. Atomic-scale 'double-slit' studies have also been considered. As an example, for aligned diatomic molecules the ionic sites may act as 'internal double-slits' for the electron wave following ionization [8-17]. Interference between partial waves then emerges in the photoelectron's energy and angular distribution, with some trace remaining even for randomly oriented targets.

The first interference experiments on single-atom photoionization was by Blondel *et al* [18], who established direct observation of the photoelectron wavefunction and its coherence, visualizing the radial nodes and allowing interference studies. Their photo-detachment microscope is not time resolved, and so doesn't reveal the dynamic characteristics of the interference phenomena. Recently, interferences in proton impact ionization of helium were measured, and reproduced theoretically using an ab-initio time-dependent model [19].

In the experiment presented here a different route is chosen to reveal information about atomic quantum interference, as in Fig.1. In the presence of a continuous-wave (CW) infrared (IR) *and* blue laser, the ground $5\,^2S_{1/2}$ state of a Rubidium atom is excited either to the $5\,^2P_{3/2}$ or $6\,^2P_{3/2}$ states using radiation at ~780.24 nm or ~420.30 nm respectively. The ionization energy of Rb is ~4.18 eV, and so the *5P* state can be ionized using blue radiation (path1), or alternatively the *6P* state ionized using infrared radiation (path2). In each case, photoelectrons emerge with *identical* energy of 0.36 eV, and are detected as a function of angle $\theta$ from the laser polarization direction.

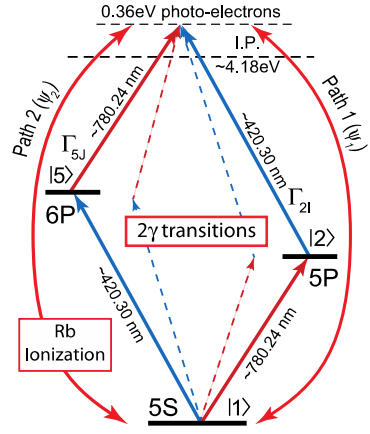

**Fig. 1** Irradiation of Rb with two independent photons generates a photoelectron passing through an intermediate state either via path1 (transition amplitude $\psi_1$) or path2 ($\psi_2$), resulting in interference related to the relative phases of $\psi_1$ and $\psi_2$. The (very weak) non-resonant 2-photon transitions are represented as dashed lines.

It is emphasized that the lasers are not phase-related and so are not mutually coherent. As a result, both pathways are activated at the same time, with no preference. The source of observed interference are hence coherences in the single atom that are reflected in the phase-relation between excitation amplitudes for the transitions in Fig.1. This is important, as interference would be less prominent for transitions that are not phase-locked, or if the phase of one intermediate state is randomized (e.g. due to coupling to a phononic or vibronic heat bath). As an example, imagine the same scheme in Fig.1 in a multilayer system, where one level belongs to one layer and the other to a second hot layer isolated from the first (e.g. by a cap layer). In this case no coherences would be generated. The same is expected if the two intermediate atomic levels have far different energies or oscillator strengths. In this sense the interference pattern unraveled here can serve as a marker for internal coherences in a sample, using non-invasively weak, readily available CW lasers. A further interesting aspect is that the role of an effective 'damping' on the interference pattern can in principle be studied in a path-selective way by detuning the respective laser, thereby altering the population time evolution of the intermediate *5P* or *6P* states.

As illustrated in Fig.1, the interaction with both lasers produces the same final photoelectron state, and so the transition amplitude describing path1 ($\psi_1$) must be added *coherently* to that of path2 ($\psi_2$) to calculate the ionization differential cross section (DCS), then given by $DCS(\theta) \propto (\psi_1 + \psi_2)(\psi_1^* + \psi_2^*)$. To access the phase relation related to the interference term, the cross sections for paths 1 and 2 are measured separately, and the incoherent sum subtracted from the coherent sum. In a simplified picture, this is analogous to closing one of the slits in Young's experiments and measuring the resulting pattern, albeit that our experiment is implemented in the frequency-wave vector space.

In the current experiments, if path2 is closed $(\psi_2 = 0)$, then $DCS_1(\theta) \propto (\psi_1)(\psi_1^*) = |\psi_1|^2$. If path1 is closed, then $DCS_2(\theta) \propto |\psi_2|^2$. When both paths are *open*:

$$DCS_{1+2}(\theta) \propto (\psi_1 + \psi_2)(\psi_1^* + \psi_2^*) = |\psi_1|^2 + |\psi_2|^2 + \underbrace{(\psi_2 \psi_1^* + \psi_1 \psi_2^*)}_{Interference\ Term}, \qquad (1)$$

the term in brackets being due to interference between wavefronts along each path. Hence:

$$DCS_{1+2}(\theta) \propto |\psi_1|^2 + |\psi_2|^2 + \underbrace{(\psi_2\psi_1^* + \psi_1\psi_2^*)}_{\text{Interference}} = DCS_1(\theta) + DCS_2(\theta) + DCS_{\text{Interf.}}(\theta)$$

$$\Rightarrow DCS_{\text{Interf.}}(\theta) \equiv \underbrace{(\psi_2\psi_1^* + \psi_1\psi_2^*)}_{\text{Interference Term}} \propto DCS_{1+2}(\theta) - (DCS_1(\theta) + DCS_2(\theta)) \tag{2}$$

Letting $\psi_1 = |a_1(\theta)|e^{j\chi_1(\theta)}$ and $\psi_2 = |a_2(\theta)|e^{j\chi_2(\theta)}$, we then have:

$$\underbrace{(\psi_2\psi_1^* + \psi_1\psi_2^*)}_{\text{Interference}} = 2|a_1||a_2|\cos(\chi_1 - \chi_2) \Rightarrow (\chi_1 - \chi_2) = \Delta\chi_{12}(\theta) = \cos^{-1}\left(\frac{DCS_{1+2}(\theta) - DCS_1(\theta) - DCS_2(\theta)}{2\sqrt{DCS_1(\theta)DCS_2(\theta)}}\right) \tag{3}$$

The relative contribution from individual path phase shifts can also be determined.

Quantum mechanically, we find the ionization process is described in our case by a two-photon matrix element including both pathways coherently (i.e., incorporating $\psi_1, \psi_2$). Mathematically the transition amplitude is given by:

$$M_0(p, \omega_{BL}, \omega_{IR}, q) = \underbrace{\sum_n \frac{1}{i}\int \frac{\langle p|\mathbf{E}_{BL}\cdot\mathbf{r}|n\rangle\langle n|\mathbf{E}_{IR}\cdot\mathbf{r}|q\rangle}{\epsilon_q + \omega_{IR} - \epsilon_n + i\varepsilon}}_{\text{Path 1}} + \underbrace{\sum_m \frac{1}{i}\int \frac{\langle p|\mathbf{E}_{IR}\cdot\mathbf{r}|m\rangle\langle m|\mathbf{E}_{BL}\cdot\mathbf{r}|q\rangle}{\epsilon_q + \omega_{BL} - \epsilon_m + i\varepsilon}}_{\text{Path 2}} \tag{4}$$

where $|q = 5s\rangle$ is the initial state, $|n\rangle, |m\rangle$ represent intermediate states and $|p = 0.36\,eV\rangle$ is the final continuum state. $\mathbf{E}_{BL} = \hat{\mathbf{e}}_{BL}E_{BL}$ and $\mathbf{E}_{IR} = \hat{\mathbf{e}}_{IR}E_{IR}$ are field amplitudes with polarization states ($\hat{\mathbf{e}}_{BL}, \hat{\mathbf{e}}_{IR}$) respectively. Energy conservation requires $\epsilon_p - \epsilon_q = \omega_{IR} + \omega_{BL}$. The first term in eqn. (4) represents creation of an electron-hole pair $(n-q)$ upon absorption of an IR photon. Interaction of the blue photon with the intermediate state $|n\rangle$ then leads to transition to the final (continuum) state $|p\rangle$. This term hence represents path1 (Fig.1). The second term describes when the IR photon is absorbed after the blue photon, resulting in a different intermediate state $|m\rangle$. Direct two-photon transitions to the continuum (also shown in Fig.1) are also modeled here, and becomes more relevant at higher intensities.

While the lasers act at the same time they are not phase-locked, so their relative phase difference $\phi_{BL-IR}$ is random. For CW lasers, the transition matrix element is $M_0(p, \omega_{BL}, \omega_{IR}, q) = e^{i\phi_{BL-IR}}\left(\sum_n a_n^{(1)} + \sum_m a_m^{(2)}\right)$, where $a_n^{(1)}\left(a_m^{(2)}\right)$ are terms contributing to the summation in eqn. (4). The cross section is proportional to $|M_0(p, \omega_{BL}, \omega_{IR}, q)|^2$ and so this random phase plays no role, as expected. Without loss of generality we hence set $\phi_{BL-IR} = 0$.

Of key importance for the interference term $DCS_{\text{Interf.}}$ are the angular momentum dependent scattering phase shifts

$\delta_l$ in the final (continuum) electron state $|p\rangle$: both pathways lead to superposition of an *s*- and *d*-partial wave that interferes with their counterparts of the other path, i.e:

$$M_0(p,\omega_{BL},\omega_{IR},q) = \underbrace{b_{s,Path1}e^{i\delta_s} + b_{d,Path1}e^{i\delta_d}}_{|a_1|e^{i\chi_1}} + \underbrace{b_{s,Path2}e^{i\delta_s} + b_{d,Path2}e^{i\delta_d}}_{|a_2|e^{i\chi_2}} \quad (5)$$

Here, $b_{l,Pathi}$ are complex coefficients depending on the dipole transitions in Eqn. (4). The interference term $DCS_{Interf.}$ is hence a result of the different ratios $b_{s,Pathi}/b_{d,Pathi}$ between *s*- and *d*-partial wave contributions associated with pathways i=1,2. Note in general $b_{l,Path1}$ and $b_{l,Path2}$ are not equal, leading to angular dependent pathway phases $\chi_1$ and $\chi_2$ in accordance with Eqn.(2).

For an appropriate numerical modelling a variant of the random phase approximation with exchange (RPAE) has been used [20]. This model accounts for electronic correlations, while the electromagnetic field interaction need only be taken to lowest order of perturbation theory for the laser parameters adopted here. The matrix elements in Eqn. (4) were thus extended when accounting for correlations between electrons and ion core, and include exchange. The photoionization DCS was then calculated in the standard way. Details of the calculations are found in the *supplementary material* [21].

In practice, Rb is more complex than shown in Fig.1, since there are two isotopes with relative abundance $^{85}Rb:^{87}Rb = 72.2\%:27.8\%$, both having hyperfine structure. The lasers adopted here have resolution $>1:10^9$ and so can excite individual hyperfine states. $^{85}Rb$ was hence excited from the $5^2S_{1/2}(F'=3)$ state to the $n^2P_{3/2}(F=4)$ states, with $n=5,6$. The Doppler profile of the atomic beam and laser power broadening influence state selectivity, and this was determined by scanning through different states while monitoring photoelectron yield. This confirmed the $F=4$ state was well resolved for the $5P$ transition, and was dominant within the $6P$ manifold.

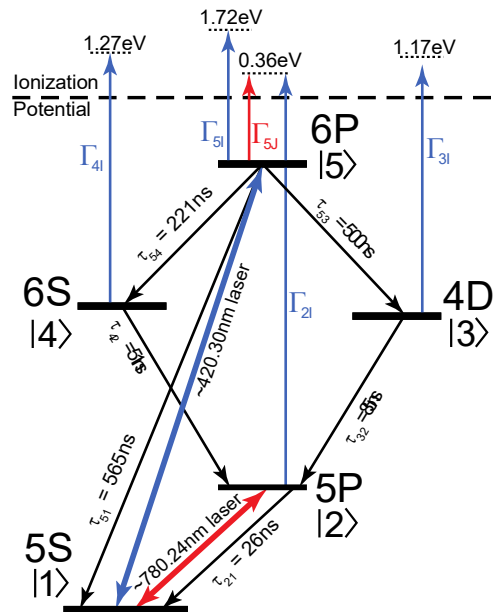

**Fig. 2** Excitation, ionization and decay processes in Rb relevant to the experiments.

A further complexity arises due to cascades from the 6P state (Fig. 2). Once the 6P state $|5\rangle$ is excited, the atom can be photo-ionized by the IR beam (rate $\Gamma_{5J}$) or by a second blue photon (rate $\Gamma_{5I}$). The atom may also relax to the 6S state $|4\rangle$ or 4D state $|3\rangle$ with lifetimes as shown. Atoms relaxing to these states decay to the 5P state $|2\rangle$, or may be photoionized by blue radiation with rates $\Gamma_{4I}$ and $\Gamma_{3I}$. Cascading into the 5P states hence add to the photoelectron yield from this state, which has rate $\Gamma_{2I}$. Atoms in the 5P state also can decay to the 5S state. Photoionization by IR light only occurs from the 6P state, producing photoelectrons with 0.36 eV energy. By contrast, the blue radiation can ionize all states, producing photoelectrons with energies as in Fig.2. The spectrometer had a resolution of 90 meV, and so easily distinguished 5P and 6P photoelectrons from 6S and 4D contributions.

Since cascades add to the 5P state yield they must be carefully considered. Fig.3 shows the pathways producing 0.36 eV photoelectrons in more detail. In Fig.3(a) the blue laser is detuned from resonance by +1500 MHz, so only the 5P state is excited. This corresponds to deactivating the second ionization pathway, since excitation to the 6P state is reduced to a negligible level. This detuning was chosen since it is mid-way between the ground hyperfine states (separation 3035 MHz). Thus, in Fig.3(a) photoelectrons can only be produced by two-photon excitation as in path1. In 3(b) the IR laser is detuned by +1500 MHz so the 5P state is not excited, and the blue beam is set resonant with the 6P state. Two contributions then occur, represented by path2 in 3(b), and also via cascades to the 5P state through 4D and 6S states. Fig.3(c) shows when both lasers are resonant. In this case contributions arise from interference (as in Fig.1), and also from cascades.

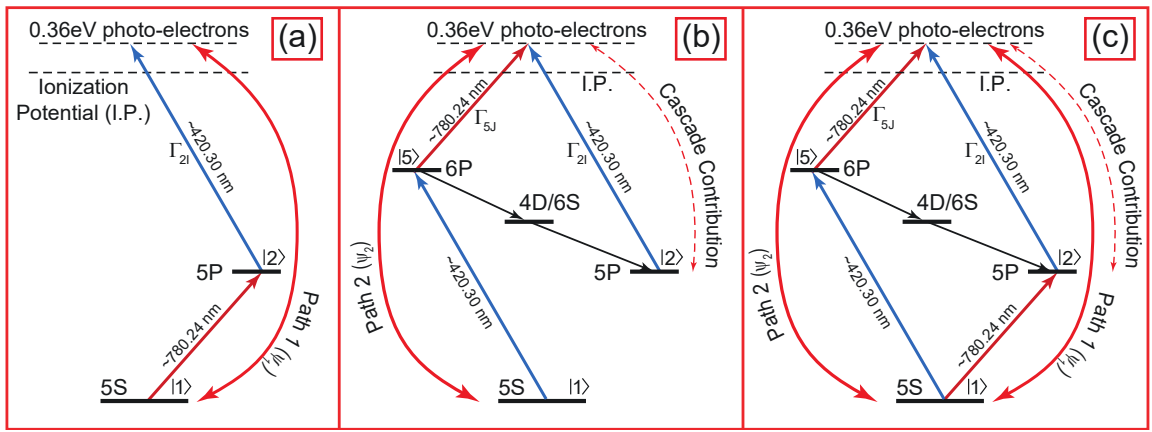

**Fig. 3** Different pathways to ionization producing photoelectrons with 0.36 eV.

It is clearly advantageous to minimize cascade contributions. To facilitate this, the long decay lifetimes through 4D and 6S states (Fig.2) were exploited to reduce cascade contributions to ~9% of the total yield. Details regarding the methods to minimize these contributions are found in [21].

To maximize contrast in any interference study, it is advantageous to ensure the amplitudes along each path closely match. This was achieved by detecting signals from each at their peak (Figs.4(a) and (b)), and then adjusting the experimental parameters so these were similar. Path1 (Fig.3(a)) produced the strongest signal, and so to balance the amplitudes the IR laser was detuned by +50 MHz to reduce the $5^2P_{3/2}(F=4)$ population prior to ionization by blue light. This method was used as the blue power could not be increased, and since detuning the IR laser was straightforward.

The experiments were hence carried out in stages. The IR laser was first detuned by +50 MHz and the blue laser detuned by 1500 MHz, so that only the $5^2P_{3/2}(F=4)$ state was excited (Fig.3(a), path1). The cross-section $DCS_1(\theta)$ was then determined. The second experiment retuned the blue laser to resonance and switched off the IR laser, so only cascade contributions $DCS_{Casc.}(\theta)$ were measured. The IR laser was then switched on and detuned by 1500 MHz to eliminate direct excitation of the $5P$ state (Fig.3(b), path2). The DCS for this process was then $DCS_2(\theta) + DCS_{Casc.}(\theta)$. The final experiment set both lasers on-resonance, with the IR laser again blue detuned to balance yields. These experiments measured $DCS_{1+2}(\theta) + DCS_{Casc.}(\theta)$ (Fig.3(c)). Since cascade contributions are incoherent, they only add to the overall yield and do not influence the interference term.

Fig.4 shows the result of these studies, the data being normalized with both lasers on resonance (Fig.4(d)). To establish the normalization accurately, the data were fitted to functions of the form $DCS_i(\theta) = \sum_{n=0}^{2} a_{ni} \cos^{2n}(\theta)$, which are symmetric around the polarization direction as required. These fits are shown in the figure. Two calculations using the RPAE model are presented, one with equal intensity pathways, and one where the paths have relative amplitudes of 1.03 and 0.97 respectively to simulate cascade effects.

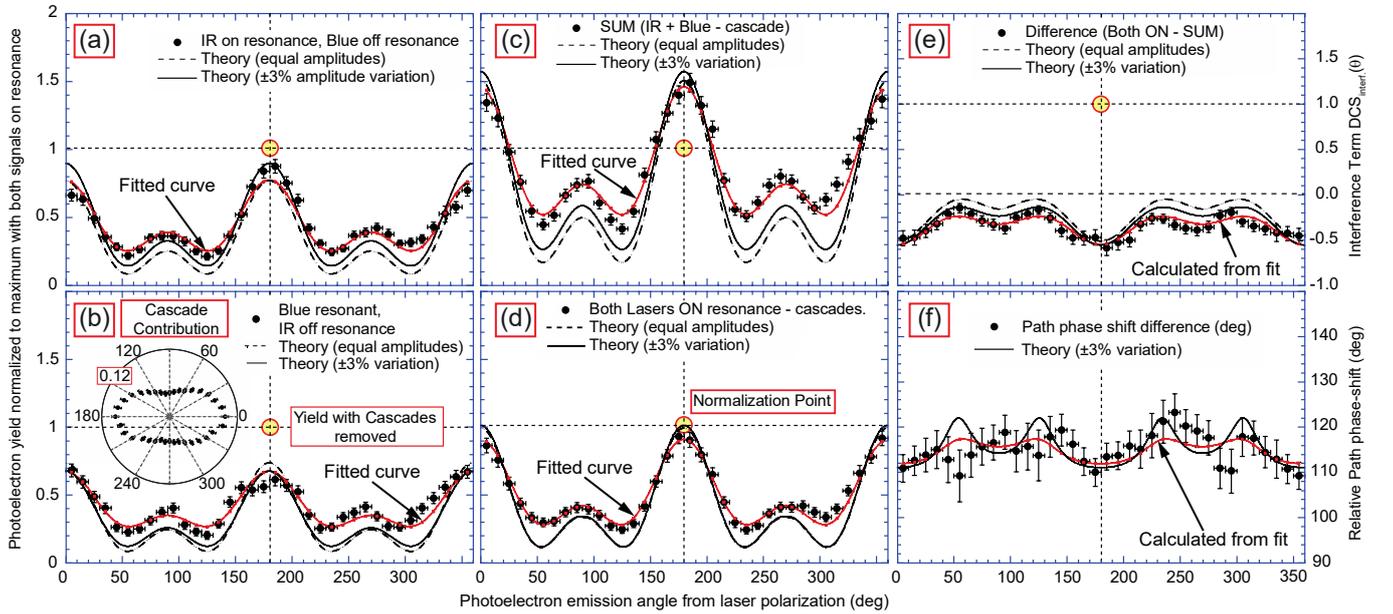

**Fig. 4** Results from experiment and theory, normalized when both lasers are on resonance. The data are compared to theory for path amplitudes equal (dashed line) and to allow for cascades (solid line). The red curves are fits to determine peak amplitudes. (a) shows results from path1. (b) shows results with only the blue laser resonant, and also shows cascade contributions (inset). The data have these contributions removed. (c) is the sum of (a) and (b), and (d) is when both lasers are resonant. (e) is the difference between (c) and (d), and so measures the interference term (Eqn.2). (f) is the relative path phase-shifts (Eqn.3), the theoretical curve showing this difference derived from individual pathways.

The cascade contributions shown as an inset to 4(b) depicts the yield when the IR laser was off. Fig.4(a) (path1) is when the blue laser was detuned, corresponding to $DCS_1(\theta)$. Fig.4(b) is the yield when the red laser was detuned

(path2), while Fig.4(c) is the sum of 4(a) and 4(b) $\left(DCS_1(\theta)+DCS_2(\theta)\right)$. Fig.4(d) shows when both lasers were resonant $\left(DCS_{1+2}(\theta)\right)$. Panel 4(e) is the difference between (d) and (c), corresponding to $DCS_{Interf.}(\theta)$. The phase shift difference between pathways calculated from Eqn.(3) is presented in 4(f). We note that a non-zero result in 4(f) at any angle is proof of the predicted interference, as follows from Eqn.(1). The red curve is from the data fit, and the theoretical curve is the calculated phase shift difference along each path, taken from the model.

Agreement between theory and experiment improves when cascades are included (by introducing imbalances between the laser field amplitudes in the calculations). The model underestimates the data around $\theta = 90°$ and 270°, however agrees well at $\theta = 0°$ and 180° (along the polarization vectors). These differences nearly cancel for the interference term $DCS_{Interf.}(\theta)$ in Fig.4(e), where the model that includes cascades agrees closely with the data. The interference term is clearly non-zero, reflecting the large difference with both lasers on (4(d)), and when individual signals are added (4(c)). This term is *negative*, so the cosine in Eqn.(3) must be negative. The relative phase difference is hence between 90° and 180°, as in Fig.4(f). The visual comparison between theory and experiment is poorer here than in figures 4(a)-(e), however the uncertainties are relatively large due to error propagation through the arccos function. To aid in comparison, the red curve in Fig.4(f) shows the phase-shift calculated from the data fits, and this shows the same trend as predicted by theory.

If the ionization pathways were incoherently related (i.e. were independent of each other as expected classically), no difference would be found between figures 4(c) and 4(d). Their subtraction would then yield $DCS_{Interf.}(\theta) \equiv 0$ at all angles, with no phase difference. This is clearly inconsistent with the data as shown in 4(e) and 4(f), and so these results clearly demonstrate the quantum nature of the two-path ionization process, and the resulting interference between different pathways.

Figures 4(e) & 4(f) show that for simultaneous (*5P/6P*) excitation the interference term is large, with an amplitude varying from 13% to 55% of the normalized signal, and a path phase difference ranging from 110° to 122°. To elucidate how sensitive these terms are to both angle and energy, they have been calculated for simultaneous excitation to the (*5P/7P*), (*5P/8P*) and *(10P/11P)* states. These calculations predict the interference amplitude and relative phase will increase as the energy gap decreases, and the angular variation will also increase. A detailed study of these effects as well as their evolution when pulsed fields are used is currently underway.

In summary, this new type of 'double-ionization path' interference allows insight to be obtained into the various facets of coherences in a sample. The experiment allows individual pathways to be controlled in a dynamic way by changing the laser parameters. These ideas can be applied to other systems, including when the final state is a highly-excited Rydberg state. This opens up possibilities for studying phase-related phenomena in Rydberg aggregates, which are currently under consideration as candidates for quantum computing.

**Acknowledgements:** JP would like to thank the University of Manchester for a doctoral training award for this work. The EPSRC UK is acknowledged for current funding through grant R120272. We would like to thank Dr Alisdair